\documentclass[aps,prd,amsmath,superscriptaddress,nobibnotes,nofootinbib,preprintnumbers,onecolumn,12pt,tightenlines]{revtex4}
\pdfoutput=1
\usepackage{graphicx}

\usepackage{graphicx}

\usepackage{ifthen}

\graphicspath{{figures/}}

\usepackage[normalem]{ulem}
\usepackage[usenames]{color}
\definecolor{DarkRed}{rgb}{0.5,0.0,0.0}
\definecolor{DarkGreen}{rgb}{0.0,0.5,0.0}
\definecolor{DarkBlue}{rgb}{0.0,0.0,0.5}
\definecolor{Brown}{cmyk}{0.0,0.8,1,0.6}
\definecolor{Magenta}{rgb}{1.0,0.0,1.0}
\definecolor{DarkMagenta}{rgb}{0.5,0.0,0.5}

\newcommand{\ifmulticol}[2]{%
  \ifthenelse{\lengthtest{1.9\columnwidth<\textwidth}}{#1}{#2}%
}

\newcommand{\gae}{%
  \ensuremath{\lower 2pt \hbox{%
    $\, \buildrel {\scriptstyle >}\over {\scriptstyle \sim}\,$}%
    }%
  }
\newcommand{\lae}{%
  \ensuremath{\lower 2pt \hbox{%
    $\, \buildrel {\scriptstyle <}\over {\scriptstyle \sim}\,$}%
    }%
  }

\newcommand{\Enr}{\ensuremath{E_{nr}}}

\newcommand{\mchi}{\ensuremath{m_{\chi}}}
\newcommand{\rhochi}{\ensuremath{\rho_{\chi}}}

\newcommand{\vesc}{\ensuremath{v_\textrm{esc}}}

\newcommand{\bv}{\ensuremath{\mathbf{v}}}  

\newcommand{\sigmaSI}{\ensuremath{\sigma_{\mathrm{SI}}}}

\newcommand{\sigmapSI}{\ensuremath{\sigma_{\mathrm{p,SI}}}}

\newcommand{\mup}{\ensuremath{\mu_{\mathrm{p}}}}

\usepackage[]{natbib}

\usepackage[hyperfootnotes=false]{hyperref}

\begin{document}


\preprint{MCTP--12-14.}


\title{New Dark Matter Detectors using Nanoscale Explosives }

\author{Alejandro Lopez-Suarez}
\email[]{aolopez@umich.edu}
\affiliation{
  Department of Physics,
 University of Michigan,
 Ann Arbor, MI 48109}

\author{Andrzej Drukier}
\email[]{adrukier@gmail.com}
\affiliation{
 Biotraces, Inc.
 Fairfax, VA}

\author{Katherine Freese}
\email[]{ktfreese@umich.edu}
\affiliation{
  Department of Physics,
 University of Michigan,
 Ann Arbor, MI 48109}

\author{Cagliyan Kurdak}
\email[]{kurdak@umich.edu}
\affiliation{
  Department of Physics,
 University of Michigan,
 Ann Arbor, MI 48109}

\author{Gregory Tarle}
\email[]{gtarle@umich.edu}
\affiliation{
  Department of Physics,
 University of Michigan,
 Ann Arbor, MI 48109}

\date{\today}



\begin{abstract} 


We present nanoscale explosives as a novel type of dark matter detector and study the ignition properties. When a Weakly Interacting Massive Particle WIMP from the Galactic Halo elastically scatters off of a nucleus in the detector,
the  small amount of energy deposited  can trigger an explosion. 
For specificity, this paper focuses on a type of two-component explosive known as a nanothermite, consisting 
of a metal and an oxide in close proximity.  When the two components interact they undergo a rapid exothermic
reaction --- an explosion.  As a specific example, we consider metal nanoparticles of 5 nm radius embedded in an oxide.  
One cell contains more than a few million nanoparticles, and
 a large number of cells adds up to a total of 1 kg detector mass.  A WIMP interacts with a metal nucleus of the nanoparticles, depositing
enough energy to initiate a reaction at the interface between the two layers.  When one nanoparticle explodes it initiates a chain reaction throughout the cell.  
 A number of possible thermite materials are studied.  Excellent background rejection can be achieved because of the nanoscale granularity of the 
 detector: whereas a WIMP will cause a single cell to explode, backgrounds will instead set off multiple cells.
  
  If the detector operates at room temperature, 
we find that  WIMPs with masses
above 100 GeV (or for some materials above 1 TeV) could be detected; they
 deposit enough energy ($>$10 keV) to cause  an explosion.  
When operating cryogenically at liquid nitrogen or liquid helium temperatures, 
the nano explosive WIMP detector can detect energy deposits as low as 0.5 keV, making the nano explosive detector
more sensitive to very light  $<$10 GeV WIMPs, better than  other dark matter detectors.

\end{abstract} 

\maketitle


\section{\label{sec:Intro} Introduction}

The majority of the mass in the Universe is known to consist of dark matter (DM) of unknown composition.  
  Identifying the nature of this dark matter is one of the  outstanding problems
in physics and astrophysics.  Leading candidates for this dark matter are
Weakly Interacting Massive Particles (WIMPs), a generic class of
particles that includes the lightest supersymmetric particle.
These particles undergo weak
interactions and their expected masses range from 1~GeV to 10~TeV.
Many WIMPs, if present in thermal equilibrium in the early
universe, annihilate with one another, leaving behind a  relic density found to be
roughly the correct value. Furthermore, recent interest in low mass WIMPs lead us to mention Asymmetric Dark Matter models, which naturally predict light WIMPs \cite{Zurek}. 

Thirty years ago, Refs. \cite{Drukier:1983gj, Goodman:1984dc} first proposed the idea of
 detecting weakly interacting particles, 
including neutrinos and WIMPs,  via coherent scattering with nuclei.  Soon after \cite{DFS} computed
detection rates in the context of a Galactic Halo of WIMPs.  This work also showed
 that the count rate in WIMP direct detection experiments will experience an
annual modulation \cite{DFS,Freese:1987wu} as a result of
the motion of the Earth around the Sun.
Then development of ultra-pure Ge detectors permitted the first limits on WIMPs  
\cite{Ahlen:1987mn}.  Since that time, a
multitude of experimental efforts to detect WIMPs has been underway, with some of them currently claiming detection. 
The basic goal of direct detection experiments is to measure the energy deposited when weakly interacting particles 
scatter off of nuclei in the detector, depositing small amounts of energy, e.g. 1-10 keV, in the nucleus.  A recent review
of the basic calculations of dark matter detection, with an emphasis on annual modulation, may be found in \cite{Freesereview}.
Numerous collaborations worldwide 
have been searching for WIMPs using a variety of techniques to detect the nuclear recoil.  

In this paper we elaborate on a novel mechanism for direct detection of WIMPs using explosives \cite{Andrzej}.  The  small amount of energy deposited
in the nucleus by the WIMP scattering event can be enough to trigger an explosion.  The registration of such an explosion then indicates that a
WIMP/nucleon scattering event took place.  In our search for appropriate explosive materials, we realized a key limitation, which we named  ``Greg's rule."  
Everything on the surface of earth, including the conventional chemical explosives, has been constantly bombarded by ionizing particles 
coming from trace amounts of naturally occurring radioactive materials and cosmic radiation. Since conventional explosives 
can be stored in large quantities for extended periods of time
(without blowing up), we may conclude that all the conventional explosives that are currently being used in 
commercial or military applications cannot be used in DM detection applications.
This does not imply that there are no explosives that can be detonated by a single highly ionizing particle. If one were to synthesize such a material it would be highly unstable and would mysteriously explode.   We need to be ``contrarians" and test such ``unsafe" explosives, which were discovered but rejected in prior R\&D. Luckily there are two 
directions to pursue.   First, the chemical explosive, nitrogen triodine (NI$_3$), has been studied and can be ignited by a single highly ionizing particle (e.g. an $\alpha$-particle) \cite{old_stuff}.  Future work on using NI$_3$ for DM detectors will be interesting. In this paper we instead study the second approach,
nanothermites.

 Thermites have been used for more than 100 years to obtain bursts of very high temperatures in small volumes, typically a few cm$^3$. Thermites are two component explosives, consisting of a metal and either an oxide or a halide.  These two components are stable when kept separated from one another; but when they are brought together they undergo a rapid exothermic
reaction --- an explosion.  The classical examples are 
\begin{eqnarray}
\text{Al}_2 + \text{Fe}_2\text{O}_3  & \rightarrow & \text{Al}_2\text{O}_3 + 2 \text{Fe} + 851.5\  {\rm kJ/mole},\label{thermite} \\
\text{Al}_2 + \text{WO}_3  &  \rightarrow & \text{Al}_2\text{O}_3 + \text{W} + 832.0\  {\rm kJ/mole}.\\
\nonumber 
\end{eqnarray}
One advantage of thermites is the impressive number of elements, which can be used.  
Classic implementation of thermites uses micron scale (1 to 10 microns) granulation, but in recent years nano-sized granules of high explosives have been increasingly used \cite{nano_thermites}. These nano-thermites make interesting
dark matter detectors.  When a WIMP strikes the metal layer, the metal may heat up sufficiently to overcome the chemical energy barrier between the metal and metal-oxide.  An explosion results.

Nanoexplosive dark matter detectors have several advantages:
\begin{enumerate}
  \item They can operate at room temperature;
  \item Low energy threshold of 0.5 keV, allowing for study of low mass $<10$ GeV WIMPs;
  \item Flexibility of materials:  One may choose from a variety of elements with high atomic mass ({\it e.g.} Tl or Ta) to maximize the spin-independent scattering rate.  Given a variety
  of materials one can also extract information about the mass and cross section of the WIMPs;
  \item One can also select materials with high nuclear spin to maximize spin-dependent interaction rate;
\item Signal is amplified by the chain reaction of explosions;
\item Excellent background rejection due to physical granularity of the detector.
Because the cells containing the nanoparticles are less than a micron in size, the detector has the resolution to differentiate between WIMP nuclear recoils, which only interact with one cell of our detectors, and other backgrounds (such as $\alpha$-particles, $\beta$-particles and $\gamma$-rays) which travel through many cells. Thus, if the background has enough energy to cause the ignition of one cell, then it would ignite multiple cells. In the section $\textbf{Backgrounds}$, the typical ranges ( $\gtrsim 10$  $\mu$m) of $\alpha$ and $\beta$ particles are shown.
\item Depending on the specifics of the detector design, the possibility of directional sensitivity with nanometer tracking; this possibility
will be studied in future papers.
    \end{enumerate}

To allow for specific calculation we study oxide-based nano-thermites, which consists of metal spheres with a radius of 5 nm embedded in an oxide. Motivated by their optical, magnetic and electronic applications, metal nanoparticles have been synthesized using both liquid and gas phase methods \cite{Oushing}\cite{Kruis}. In situations where the metal nanoparticles are susceptible to oxidation, the nanoparticles can be coated by a thin layer of an inert metal \cite{O'Conner}. To form a nano-thermite the metal particles must be mixed by an appropriate gel of oxide \cite{Nano-wire}. Alternatively, the oxide can be replaced by an appropriate halide \cite{Andrzej}.

Enough energy deposit in the metal sphere heats it up to the point where there is an explosion beginning at the interface of the two
materials at the edge of the nanoparticle.
As a specific design, we imagine constructing a ``cell" which consists of $\sim 10^6$ metal nanoparticles embedded in an oxide.  A full detector will need
many of these cells; e.g. to obtain 1 kg of target material (the metal) there will be $\sim 10^{14}$ cells. A WIMP hitting the target will
cause only one of these cells to explode.

 \begin{figure}[h!]

\includegraphics[width=1.1\linewidth]{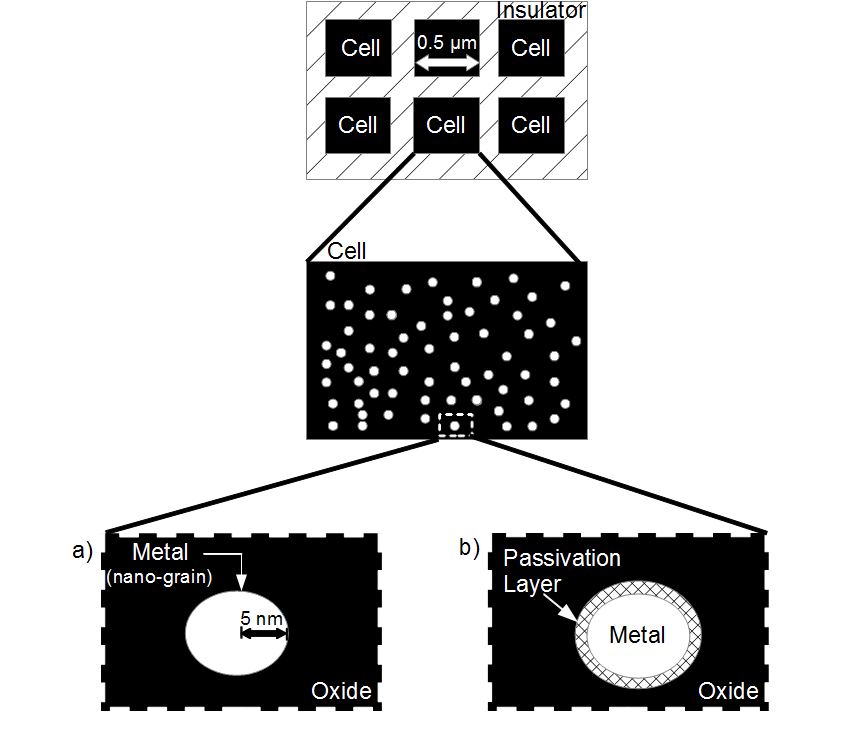}
    \caption{This figure depicts a schematic view of the nano-thermite detector studied. An array of cells of length $0.5 \ \mu$m is embedded into an insulator, which thermally decouples the cells from each other. Each cell contains more than a few million metal nanoparticles embedded into a metal-oxide. Two different images are depicted at the bottom of the figure: (a )shows the design model used for all calculations, and (b) represents a more realistic depiction of the nano-thermite detector. The dissimilarity between both images is the addition of a passivation layer in image (b). A passivation layer is a metal-oxide coating placed around the nanoparticle in order to prevent oxygen molecules interacting with the metal. An oxidized metal will not react chemically with a metal-oxide, since it is no longer favorable to gain oxygen atoms. Thus, an oxidized metal will not produce a thermite reaction. The passivation layer covering the metal nanoparticle would be required in the synthesis of the detector; since it would prevent oxidation of the metal nanoparticle during construction of the detector (i.e. before embedding the nanoparticle into the cell). As well, in some differing implementations, the metal-oxide of the cell could be comprised of mixed nano-wires \cite{Nano-wire}, which would produce a larger temperature increase due to a higher effective thermal resistance between the oxide and the metal. Image (a) represents a simplified design model, which enabled analytic results in later sections.}
    \label{detectorfinal}
\end{figure}

More precisely, when a WIMP elastically
scatters with a metal nucleus and deposits energy to the metal,
 then that energy is converted into a temperature
increase. If the temperature increase is big enough to overcome the
potential barrier of the thermite reaction, then the metal will
react with the surrounding oxidizer exothermically. In the design using metal nanoparticles, after the first
thermite reaction of one nanoparticle occurs, the exothermic heat produced
by the thermite will heat up the other metal nanoparticles within the 0.5
$\mu$m cell; thus creating a chain reaction which amplifies the signal
to a measurable effect. Utilizing Eq \ref{thermite} as an example, the amplification factor for the signal is on the order of $10^4$-$10^5$. The detection of the cell explosions could
be made by sensitive microphones or spectroscopic studies of the debris. 

Figure \ref{detectorfinal} shows a schematic representation of the nano-thermite detector studied in this paper. On top, the first picture of Figure \ref{detectorfinal} shows an array of cells embedded into an insulating material. The insulator is used to thermally decouple the cells; so that the reaction within a cell does not cause the explosion of neighboring cells. The length of each cell is taken to be $0.5 \ \mu$m. The spatial scale of the cells enable us to distinguish background from WIMP/nucleus collisions. Backgrounds composed of $\alpha$, $\beta$ and $\gamma$ particles will traverse multiple cells; whereas a recoiled ion from a WIMP/nucleus collision will only interact with a single cell. The middle picture in the figure is a magnified view of an individual cell. Inside each cell there will be more than a few million nanoparticles. The nanoparticles, represented by the white circles, are embedded into the metal-oxide, shown as the black background. Finally, the bottom pictures of Figure \ref{detectorfinal} depicts an enlarged section of the cell surrounding a single nanoparticle of radius $5$ nm. There are two pictures at the bottom. Image (a) shows the simplified model used to make all the calculations in the sections $\mathbf{Temperature\  Increase}$ and $\mathbf{Results}$. In contrast, image (b) depicts a more realistic design for the nano-thermite detector. A thin passivation layer is placed around the metal to prevent oxidation of the metal during the construction of the detector (i.e. before embedding the metal nanoparticle into the cell). The passivation layer is a metal-oxide coating placed around the nanoparticle in order to prevent oxygen molecules interacting with the metal. An oxidized metal will not react chemically with a metal-oxide, since it is no longer favorable to gain oxygen atoms. Thus, an oxidized metal will not produce a thermite reaction. However, the passivating barrier is lowered if the metal nanoparticle or the passivation layer melts due to the temperature increase. In a realistic scenario, the synthesis of the nanoparticle embedded into the oxide would require a passivation layer. It should be noted that the addition of an extra layer between the metal and the oxide of the cell would produce an additional thermal resistance at the interfaces. This thermal resistance would cause the metal to hold in heat; and thus, increase the temperature increase yield after a WIMP/nucleus collision, when compared to the results presented in this paper. As well, in some differing implementations, the metal-oxide of the cell could be comprised of mixed nano-wires \cite{Nano-wire}, which would produce a larger temperature increase due to a higher effective thermal resistance between the oxide and the metal. As explained in the section $\mathbf{Temperature\  Increase}$, the temperature increase is calculated utilizing the design model of image (a) (i.e. no passivation layer) and zero thermal resistance between the oxide and the metal nanoparticle. Thus, our calculations are conservative and underestimate the temperature increase sourced by an elastic collision between a WIMP and a metal nucleus.  

 More generally, many other detector designs may be possible, such as two parallel layers of the two components. 
This latter design would allow determination of the direction from which the WIMP came, as only WIMPs headed first into the metal (rather than first into the oxide)
would initiate an explosion. 

The goal of this paper is to study the ignition of the explosion when a WIMP hits the metal nanoparticle.  A parallel paper \cite{Andrzej} studies
the nano boom dark matter detectors more generally, including methods of detection and readout of the explosion; alternate  explosives other than thermites; 
and other aspects of the problem. 

In this paper we begin by reviewing the relevant particle and astrophysics of direct detection, and then turn to the viability of a nanothermite detector for WIMPs.
For our calculations we consider WIMP masses of $m_{\chi}=10,100\ \text{and\ }1000$ GeV.

\section{\label{sec:DMDetection} Dark Matter Detection}

WIMP direct detection experiments seek to measure the energy deposited
when a WIMP interacts with a nucleus in a detector.
If a WIMP of mass $\mchi$ scatters elastically from a nucleus of mass
$M$, it will deposit a recoil energy 
\begin{equation}
\label{eq:energy_deposit}
\Enr = (\mu^2 v^2/M)(1-\cos\theta) ,
\end{equation}
where $\mu \equiv \mchi M/ (\mchi + M)$ is the reduced mass of the
WIMP-nucleus system, $v$ is the speed of the WIMP relative to the
nucleus, and $\theta$ is the scattering angle in the center of mass frame.
The typical energy transferred to the nucleus in a scattering event is from 0.1 to 50 keV, depending on the WIMP mass and the detector material. 
Typical count rates in detectors are less than 1 count per kg of detector per day. 
Reviews of the dark matter scattering process and direct
detection can be found in Refs.~\cite{Primack:1988zm,Smith:1988kw,
Lewin:1995rx,Jungman:1995df,Bertone:2004pz}.
Over the past twenty five years a variety of designs have been developed to detect WIMPs. 
They include detectors that measure scintillation; ionization; and dilution-refrigerator based calorimeters which measure the total energy deposed  by means of 
a phonon spectrum.  Current detector masses range in size up to 100 kg.  The upcoming  generation of detectors will reach one tonne.

A major concern in all WIMP detectors is backgrounds. To eliminate spurious events from CR, the detectors must be placed deep underground 
( $>$ 2,000 m of water equivalent). Yet radioactive backgrounds remain and must be eliminated. Thus the experimental determination of 
annual and/or diurnal modulation is a crucial test of the WIMP origin of any events observed in the detector, as most backgrounds should not
exhibit the same time dependence.
     
\medskip\noindent
{\bf Particle Physics:  WIMP/nucleus cross sections:}

For a supersymmetric (SUSY) neutralino and many other WIMP candidates, the dominant
WIMP-quark couplings in direct detection experiments are the scalar and
axial-vector couplings, which give rise to spin-independent (SI)
and spin-dependent (SD) cross-sections for elastic scattering of a
WIMP with a nucleus, respectively.
 SI scattering is typically taken to be
\begin{equation} \label{eqn:sigmaSI2}
  \sigmaSI = \frac{\mu^2}{\mup^2} A^2 \, \sigmapSI \, ,
\end{equation}
where $A$ is the atomic mass of the nucleus, $\mup$ is the WIMP-proton reduced mass and $\sigmapSI $ is the SI scattering cross section of WIMPs with protons.
For large momentum transfer, this relation is multiplied by a form factor correction to account for the
sensitivity to the spatial structure of the nucleus.
Since the SI cross-section grows rapidly with nuclear mass, direct detection experiments often use heavy
nuclei to increase their sensitivity to WIMP scattering.

Spin-dependent (SD) WIMP-nucleus interactions depend on the spin of the nucleus.  Most nuclei have
equal numbers of neutrons and protons so that there is no SD contribution;  specific nuclei
must be chosen in experiments to search for nonzero SD couplings.  
SD scattering is often of lesser significance than SI scattering in direct detection experiments
 for the heavy elements used in most detectors due to the extra $A^2$ coherence factor
 in the cross section.

\medskip\noindent
{\bf Astrophysics: Velocity Structure of the Galactic Halo:}

The velocity distribution $f(\bv)$ of dark matter particles in the
Galactic Halo is crucial to their signals in dark matter detectors (as first stressed by \cite{DFS}).
The dark matter halo in the local neighbourhood is likely to be
composed mainly of a smooth, well mixed (virialised) component with an
average density  $\rhochi \approx 0.4$~GeV/cm$^3$. The simplest model
of this smooth component is the Standard Halo Model (SHM),
 a spherically symmetric nonrotating isothermal sphere with an isotropic, Maxwellian velocity
distribution characterized by an rms velocity dispersion $\sigma_v \sim 290$ km/sec; the distribution 
 is truncated at  escape velocity $\vesc \sim$ 550 km/sec in the Galactic rest frame, or $\vesc \sim$ 750 km/sec in the laboratory
 frame (the frame of the Earth) where the measurements are made.   The resultant count rates in direct
 detection experiments due to the SHM were first discussed in \cite{DFS}. 

A key issue for dark matter detectors using nano-explosives is the question of whether or not
a WIMP deposits enough energy in the detector  to initiate an explosion.  
The WIMP velocity plays a key role, as can be seen in Eq. (\ref{eq:energy_deposit}).  We will
study two different WIMP velocities.  First, we take the typical WIMP speed, which is roughly 300 km/sec in the lab frame.  However,
it is possible that, although the typical WIMP is not sufficient to set off an explosion, there are plenty of high velocity WIMPs on the tail
of the distribution that do cause explosions.  In fact many existing dark matter detectors (such as LUX, XENON, and CDMS) 
rely on this tail in obtaining results for low
mass WIMPs near the threshold of their sensitivities.  Thus, as our second case, we will consider fast WIMPs with speeds of 700 km/sec 
in the lab frame. Herein, we study both cases for oxide-based nano-thermites; and aim to extend the study both cases for halide-based nano-thermites in future work.

A related question is the value of the energy recoil, which depends on the scattering angle (see Eq. (\ref{eq:energy_deposit})).
The  maximum recoil energy takes place for forward scattering with an  angle of
$\theta = \pi$, i.e. 
\begin{equation}
\label{eq:Emax}
E_{max} =  2 \mu^2 v^2/M.
\end{equation}
 In some cases a more typical value of the energy recoil is half of the maximum,
 but an accurate calculation including form factors for the detailed interactions with the nucleons inside the nucleus would be required
and is the subject of future work.  For our case of 300 km/sec WIMP speed, we assume a characteristic energy of 1/2 the maximum energy recoil; whereas for the fast WIMPs with
700 km/sec, we assume the maximum energy recoil.  In the future it would be interesting to integrate over the entire WIMP velocity
distribution, with results  dependent on the 
detector material of choice as well as WIMP mass. 

\medskip\noindent
{\bf Current Experimental Status for Direct Detection:}

In the past decade, a host of direct detection experiments using a variety of different detector materials and designs
 have reported unexplained nuclear recoil signals which could be due to WIMPs. 
Detection of annual modulation has now been claimed by the DAMA and, more
recently, CoGenT experiments.
The Italian Dark Matter Experiment, or DAMA \cite{DAMA}, consists of 250 kg of radio pure NaI scintillator situated in the Gran Sasso Tunnel underneath
the Apennine Mountains near Rome, and became
the first direct detection experiment to observe a
positive signal.  
The group now has accumulated 1 ton-yr of data over the past decade and
  finds an 8.9 $\sigma$ annual modulation with the correct phase and spectrum to be consistent with a dark matter signal.
 Recently CoGeNT \cite{COGENT}, consisting of Germanium, also
claim to see annual modulation of the signal with the correct phase to be consistent with WIMPs, and together with a third
CRESST-II \cite{CRESST} experiment, could be seeing $\sim$10 GeV WIMPs.  The CDMS experiment also
has seen a few unexplained events in their Silicon detectors at low energies \cite{Agnese:2013rvf}, that might be compatible with low mass WIMPs.
 However, other experiments, notably CDMS-Ge \cite{Ahmed:2012vq, Ahmed:2010wy}, SuperCDMS \cite{SuperCDMS}, XENON \cite{Angle:2011th} and LUX \cite{Akerib:2013tjd},
have null results that  conflict with these positive signals and may rule them out.   
Many direct detection experiments are either currently running or gearing up to do so, and we can expect more data soon. 
 
 The COUPP \cite{COUPP1}\cite{COUPP2}, PICASSO \cite{PICASSO},  SIMPLE \cite{SIMPLE}, and PICO \cite{PICO} (a new collaboration between the COUPP and PICASSO teams) experiments are the most similar to the nano explosives proposed here.  
 PICASSO and SIMPLE use superheated droplet detectors and COUPP uses bubble chambers.  These detectors  go through a phase transition when hit by a WIMP; 
 easily visible bubbles form in the bubble chambers.  These detectors operate at room temperature and are sensitive to low mass WIMPs, down to about 15 GeV. The new PICO collaboration aims
 for a 3 keV energy threshold, allowing it to study WIMPs of even lower masses; their limitations are due to gamma-ray backgrounds.  
 One of our goals is to explore new materials and designs that allow the detector to have even lower threshold and sensitivity
 to even lighter WIMPs.  Because our targets are nm in size, we have the resolution
 to differentiate between WIMP nuclear recoils, which only interact with one cell of our detectors, and other backgrounds (such as $\alpha$s and $\gamma$s) which
 travel through many cells and cause multiple ignitions. 
 
In the past few years the cross-sections that have been reached by detectors have improved by three orders of magnitude; over the next few years
another two orders of magnitude should be reached.  The next generation of detectors being built will be one tonne in mass or directional.  
A review of the theory and experimental status of dark matter detection can be found in \cite{Freesereview}.  One of the goals of using nano explosive DM 
detectors is to design low threshold detectors that can test the light mass $<$10 GeV hypothesis.

\medskip\noindent
\section{\label{sec:Intro} Basic Idea of Nanothermites as WIMP detectors}

For a given energy deposit by a WIMP (given in Eq. (\ref{eq:energy_deposit}), the amount that the nanoparticle heats up is determined by
\begin{equation}
\label{eq.heat_capacity}
\Delta T = \Delta E/C_{n}
\end{equation}
where the heat capacity of the nanoparticle $C_{n}$ depends on the material as well as on the size of the nanoparticle.  

It is instructive to see specific examples of Eq [\ref{eq.heat_capacity}] in order to get an idea of the amount of temperatures roughly expected to be produced inside the nanoparticle. The specific heat per volume of Al is $c^{Al}=1.5 \times 10^{-5}$ keV/nm$^3$.  The heat capacity for an aluminum nanoparticle of radius $R=1 $\ nm is $C^{Al}_{n}=c^{Al} \frac{4\pi R^3}{3}=6.3 \times 10^{-5}$ keV/K. Thus, the expected temperature increase for an aluminum nanoparticle with an energy of $1$ keV deposited by a WIMP interacting with an aluminum nucleus is $\Delta T=\frac{1keV}{C^{Al}_n}=1.59\times 10^4$ K. The ignition temperature for a micron-size $\text{Al}/\text{Fe}_2\text{O}_3$ thermite necessary for the reaction to begin is roughly $1,000$ K, which implies that a nanoparticle with radius ($R_n = 1$ nm) could produce high enough temperatures for the thermite reaction to begin and produce an explosion. The low values of the heat capacity for metal nanoparticles allows for a very large temperature increase. This is one of the main reasons for pursuing nano-thermites as a possible WIMP detector. In this paper, we aim to improve our previous calculation with Eq [\ref{eq.heat_capacity}] in order to establish if the nanoparticle could work as a WIMP detector or not. There are two issues that need to be considered.

 The first complication arises from the fact that the metal nanoparticle will not retain the temperature increase for an infinite amount of time and will dissipate heat following the dynamics of the heat transfer equation. The question of whether the thermite reaction will begin or not needs to be addressed by studying the heat transfer equation in order to establish the amount of time the nanoparticle stays heated.  Then the Arrhenius Equation approximates the reaction rate given a specific chemical reaction, size of the nanoparticle and temperature. This reaction rate is  multiplied by the timescale determined from the heat
transfer equation to estimate the probability of a nanoparticle exploding. The Arrhenius equation allows us to estimate the quantum efficiency of detonation, the probability that a detonation would occur following a nucleus recoil event.

The second complication considered is the distance traveled by the recoiling nucleus.  It is expected that sometimes the recoiling nucleus will escape the metal nanoparticle after interacting with the WIMP. The escaping nucleus will not depose all of its energy into the nanoparticle. The range at which the recoiling nucleus stops needs to be understood in order to approximate correctly the amount of energy deposed into the metal nanoparticle. Calculations to address these two complications;
utilizing the heat transfer equation, the reaction rate and stopping distance for the recoiling nucleus in order to establish the viability of nano-thermites as WIMP detectors; will be done in later sections below. 

We may ask what temperature increase is required to initiate an explosion.  Here we describe the basic idea, and continue  in detail in the next section.
We treat the system as a phase transition with a barrier that 
must be overcome in order for the thermite reaction to take place.  
The ability to use nano-thermites for dark matter detection is pulled in two contrary directions.  On the one hand,  the material must
be chosen so that it does not spontaneously explode due to thermal fluctuations.  We may require no spontaneous explosion for at least one year; this
requirement determines the barrier height for the chemical reaction.  The required barrier height can be quite significant and is particularly
restrictive for a detector operating at room temperature.
Yet, on the other hand, we would
like the most sensitive possible detectors to incoming WIMPs.
Ideally the materials with the smallest required temperature 
increase (due to a WIMP hit) would detect the most WIMPs.  We've seen that typical WIMP interactions deposit 0.1-50 keV's of energy.
Given the barrier height required to avoid spontaneous combustion,
temperature increases of $>$1000 K are required for the thermite reaction to take place at room temperature.  We will see that heavy  WIMPs with masses
above 100 GeV (or for some materials above 1 TeV) 
are able to deposit enough energy ($>$10 keV) to cause ignition at room temperature for the detector geometry considered in this paper (other more favorable geometries
are considered in the companion paper \cite{Andrzej}.

It is also possible to operate at cryogenic temperatures, such as 77 K using liquid nitrogen or 4.2 K using liquid helium as coolants.  
At these lower temperatures, the thermal fluctuations that produce spontaneous ignition are less effective, and we may choose a material with a lower barrier
height for the phase transition without spontaneous detonation.  In this case the detector can react to lower temperature increases $\sim$ 50 K, corresponding 
to lower energy thresholds, for some materials as low as $\sim$ 0.5 keV. Some of the explosives that are designed to operate at 4.2 K may not be stable at room temperature. To realize such explosives we need to develop cryogenic methods of mixing metal nanoparticles and oxides. When operating cryogenically, 
the nano explosive WIMP detector can detect $<$10 GeV WIMPs better than any 
previous dark matter detector. 

We can see the advantages of using nano-thermites, rather than larger micron-sized ones:
\begin{enumerate}
  \item Only when the objects are nano-sized is their specific heat small enough to allow operating at room temperature.  
Smaller detector elements have smaller $C_{n}$ and thus, for the same energy deposited 
by a WIMP, larger temperature increase (see Eq.(\ref{eq.heat_capacity})).  
For nano-sized thermites the temperature increase due to WIMP interactions is large enough to cause an explosion; whereas for micron-sized
thermites the temperature increase would be too low.
    \item Low energy threshold of 0.5 keV, allowing for study of low mass $<10$GeV WIMPs, can be achieved when:
    \begin{enumerate}
     \item operating at cryogenic temperatures;
     \item using chemically active metals with low melting temperatures such as: gallium, rubidium, caesium, indium, tin, lead or bismuth;  
     \item employing F-based (fluorine-based) nano-thermites (e.g. Al $+$ WF$_6$);
     \item considering more advance models other than nanoparticles embedded in an oxide. 
     \end{enumerate}
     The possibility of reaching lower energy thresholds through the implementation of cryogenic temperatures will be further discussed in this paper. A detailed discussion for the design concepts (b)-(d) is beyond the scope of this paper and will be discussed in forthcoming papers.  
    \item   Only when the detector elements have a  size smaller than the track range of background can one use new methods of background rejection. The track lengths of recoiling nuclei are $\sim 50$ nm, which is much smaller than the typical range of an $\alpha$ particle (approximately $10\ \mu$m) emitted by radioactive decay. In nanothermites, a WIMP will make one and only one cell explode, while other background particles will cause many cells to explode. In contrast, currently employed detectors with a physical granularity greater than approximately $10\ \mu$m will not be able to make this differentiation.  
    \item    Directional detectors are possible, not with spherical nanoparticles, but with asymmetric detector designs that may be the study of future work.  Here the goal is
    to obtain the direction that the WIMP came from; for the case of forward scattering this is the same as the direction of the nuclear recoil.  Directionality would
    prove WIMP detection with much less statistics, with only $\sim$ 100 WIMPs required; in addition one would learn about the structure of the dark matter Halo.
    For directional sensitivity, the detector resolution must be smaller than the size of the track range of recoiling nuclei in order to measure the particles' track.  
    If the resolution is micron-sized, while the track length is nanometers, then of course the track will be impossible to follow.  Thus nano-scale detectors in principle
    have the capability to obtain nanometer track  resolution.  The spherical nanoparticles studied in this paper do not have directional sensitivity but other designs may.
    \end{enumerate}
    
 
 
 

\section{Activation Energy and Ignition Temperature}

We take  the rate of the thermite
reaction $\Gamma(T)$ at temperature $T$ to be represented by the Arrhenius Equation
\begin{equation}
\Gamma(T)=A N e^{-\frac{T_{a}}{T}},\label{eq:arrhenius}
\end{equation}
where $N$ is the number of interacting sites where a metal nucleus 
could chemically interact with the oxide, and $A$ is the Arrhenius prefactor which
is unique for each reaction but can be approximated by the
vibrational frequency at the interface of the reactants. The vibrational frequency of a crystal is roughly $10^{13}$ Hz. A survey of published values for the Arrhenius prefactor $A$ experimentally measured for solid decomposition reactions showed a slight predominace of $A$ values between $10^{11}$ Hz - $10^{13}$ Hz\cite{Galwey1,Galwey2}. In this paper, we will approximate $A \cong 10^{13}$ Hz, but wish to note that differences in the Arrhenius prefactor of order $10^2$ will not change our future results of the desired ignition and activation temperatures dramatically (less than an order of magnitude) due to the dominance of the exponential term. 
The Arrhenius equation Eq \ref{eq:arrhenius} will allow us to calculate the probability of ignition of the nano-thermite following a transient temperature increase.

Now we turn to computing two important temperatures:
one characterizing the barrier height of the phase transition 
and the other, the ignition temperature needed for a thermite reaction to take place.  
We will find that each of these temperatures must exceed a minimal value:
the temperature characterizing the barrier height must be high enough to prevent spontaneous
explosion even when no WIMP has hit the detector, and the ignition temperature $T_{n}$ sets a minimum
value needed in order for an explosion due to a WIMP interaction to take place.   
These two conditions are important considerations when choosing a metal and oxide to make up the
oxide-based thermite we will use in DM detectors.

{\it The first condition:}
The barrier height is characterized by the ``activation energy" \\
 $E_{a}\equiv k_{b}T_{a}$ (where $k_{b}$
is the Boltzmann constant) and its corresponding ``activation temperature"  $T_{a}$.
As mentioned above, the barrier height is determined by requiring
the thermite to be stable at room temperature $T_R = 300$ K to thermal fluctuations for at least one year, i.e.,
we require $\Gamma(T_R) \times 1\  {\rm year} < 1$. Even though the latter requirement is a conservative estimate, we will later show that a nano-thermite dark matter detector could meet such a constraining stipulation.
In adapting Eq.(\ref{eq:arrhenius}) to the requirement $\Gamma(T_R) \times 1\  {\rm year} < 1$ and computing the prefactor $N$, we must add up all
 metal nuclei that could chemically interact with the metal-oxide
and possibly produce an explosion within
the entire detector.

As mentioned previously, each cell has a radius $R\sim500$ nm. Since the typical size for the lattice constant $L$, which measures the separation between nuclei in the metal, is a few angstroms ($L\sim 5\  \text{\AA}$), 
the number of sites per nanoparticle of
radius $R_{n}=5$nm is  $N_{n}=4\pi \left( \frac{ R_{n}}{L}\right)^2=1.3\times 10^{3}.$
Thus, the total number of interacting sites $N_{det}$ found in
a detector of total mass $M_{det}$ is given by
$N_{det}  =  \frac{M_{det}}{\rho_{n}V_{n}}N_{n}$
where $\rho_{n}$ and $V_{n}$ are the density and volume of
a single nanoparticle. To obtain a numerical value we will take $M_{det} = 1$ kg.
To be as conservative as possible, we will take the lowest metal density for all
the elements we consider:  Aluminum.  Then we find $N_{det}  \cong  9.2 \times10^{23}$ and set $N=N_{det}$ in Eq.(\ref{eq:arrhenius}).
Now we can impose the condition of having a stable detector
 at room temperature that will not have any thermite
reaction in the absence of a dark matter interaction for a running
time of $1$ yr:
\begin{eqnarray}
\Gamma_{det}(T_{R})\text{\ensuremath{\times}1 yr}= & Ae^{-\frac{T_{a}}{T_{R}}}N_{det}(1\ \text{yr})\leq & 1\nonumber \\
 & T_{a}\geq 3.1 \times 10^4\ \text{K}\label{eq:Ta}\\
\nonumber 
\end{eqnarray}
 Thus, in searching through all possible thermite elements for possible use as DM detectors, we should choose
 those with thermite activation temperature $T_{a}$ greater than $3.1\times 10^4$ K. We note that because of additional methods of background rejections such as annual modulation, we could select a detector to be less stable to thermal fluctuations (e.g. $\Gamma(T_R) \times 1\  {\rm day} < 1$). Nevertheless, our previous result is very robust because of
the exponential nature of the Arrhenius equation (i.e. the same calculation of the activation temperature
for any varying conditions under the assumed simple nanoparticle/oxide model will not vary much from this result Eq$[\ref{eq:Ta}]$).

{\it The second condition:}
Now we can proceed to find the ignition temperature $T_{n}$
needed for a signal to be seen in our detector.   We want to find the minimal
temperature increase required by the energy deposited in a WIMP interaction
that can lead to an explosion.  

When a WIMP hits a nucleus inside a metal nanoparticle with radius $R_n=5$ nm, the nucleus typically traverses the entire nanoparticle
(or even somewhat farther).  We may take the entire nanoparticle to be heated by some temperature $T_n$. 
However, this value of the temperature does not last more than nanoseconds. In the next section we will 
solve the heat transfer equation to estimate the diffusion timescale of the heat out of the nanoparticle. 
Shortly after a characteristic timescale known as the ``conduction time" $t_{c}$, the metal nanoparticle is no
longer hot enough to induce ignition.  Thus the appropriate timescale with which to multiply the rate in Eq.(\ref{eq:arrhenius}) is this
conduction time.  To successfully have an exploding nanoparticle operating at some temperature $T_n$, we thus require
$\Gamma(T_{n}) t_{c} > 1$.  
We will see below that the conduction time is given by $t_{c}=\frac{R_{n}^{2}}{\alpha}$, where $R_{n}$ is
the radius of the metal nanoparticle and $\alpha$ is the thermal diffusivity
of the metal.  The biggest metal
thermal diffusivity studied will be on the order of $\alpha \sim10^{-4} $ m$^{2}$s$^{-1}$;
thus the shortest time-scale considered will be $t_{c}^{max}\sim2.5\times10^{-13}$ s. However, if the passivation layer is introduced between the metal nanoparticle and oxide, then the conduction time may be order of magnitudes longer. In our detector model, we are interested in the explosion of a single metal nanoparticle embedded in an oxide. Thus, we take $N$ to be the  number of sites on a single nanoparticle, $N=N_{n}$.

The ignition temperature $T_{n}$
is now given by considering the following inequality:
\begin{eqnarray}
\Gamma_{nano}(T_{n})t_{c}^{max}= & Ae^{-\frac{T_{a}}{T_{n}}}N_{n}t_{c}^{max}\geq & 1\nonumber \\
 & T_{n}\geq 3.8 \times 10^3\ \text{K}\label{eq:Ti}\\
\nonumber 
\end{eqnarray}
The last inequality follows from taking the lowest value of the activation energy allowed by Eq.(\ref{eq:Ta}),
$T_{a}=3.1\times 10^4$ K. Thus, in order to have a detector that can (a) run for
a full year without any spontaneous thermite reactions and (b) detect
signals when a dark matter particle interacts with the nanoparticle,
the ideal thermite reaction for the detector is one with an activation
temperature $T_{a}\geq 3.1\times 10^4$ K and an ignition temperature $T_{n}\geq 3.8\times 10^3$ K. The value for the ignition temperature needed for the nano-thermite to ignite is different from values quoted in literature due to the difference in time and size scale. In order to find the ignition temperature of a given thermite, experimentalist usually heat the bulk of the materials for prolonged periods of times when compared to the conduction time (on the order of seconds-minutes). As an example, if we substitute $t_c^{max}\rightarrow 1\ \text{sec}$, then the resulting ignition temperature $T'_n$ is closer to experimental results for thermite ignition temperatures: $T'_n=836$ K. For this reason, the minimum temperature needed  for a single nano-particle to ignite due to a recoiling metal nucleus under our assumed simple model is higher than expected.

Note that the ignition temperature is also a function of the activation
energy, so that an increase in the activation energy increases the
ignition temperature. As well, the activation and ignition temperatures were calculated after alleviating the condition of having no signals sourced by thermal fluctuations for a year ($\it{the\ first\  condition}$) to 1 per day (1 per hour): $T_a = 2.9 \times 10^4$ K ($2.8 \times 10^4$ K) and $T_{n} = 3.6 \times 10^3$ K ($3.5 \times 10^3$ K). This calculation was done in order to gain perspective under our assumed simple model into how the activation and ignition temperatures change as a function of the stability timescale of the detector. 

It should also be mentioned that the
ignition temperature might be higher than the melting temperature
of the metal and/or oxide. The change in phase of the metal would
change the previous calculation by increasing the amount of sites
able to interact chemically; and thus, lowering the ignition temperature. In order to be consistent and conservative in our calculation, we will adopt $T_{n}=3.8\times 10^3 $ K as our ignition temperature in all future comparisons. Nevertheless, it is possible to have a different detector design that lowers the ignition temperature needed without sacrificing the stability of the nanothermite detector to thermal fluctuations. Our previous calculations have been done under the assumption that the explosion of one nanoparticle of radius $R_n = 5$ nm initiates the chain reaction throughout the cell, which is interpreted as a signal. However, it is possible to conceive a detector in which two or more smaller metal nanoparticles (e.g. $R_n = 2$ nm) need to ignite in order to initiate the chain reaction throughout the cell. This multi-particle ignition mechanism suppresses activation of the nano-thermite cell due to thermal fluctuations and permits a lower ignition temperature (i.e. energy threshold). The specific design aspects for the multi-particle ignition detector is currently being pursued, but is beyond the scope of this paper and will be discussed in future work. There are many different implementations of the nano-thermite detector that are worth pursuing in more detail. In this paper we concentrate on the simple model of a metal nano-particle embedded in an oxide. The minimum ignition temperature adopted for future comparison will be the conservative estimate of $T_{n}=3.8 \times 10^3$ K.   

Now that we have calculated the ignition temperature needed, in the next section we will
study  which metals have the necessary thermal
and physical properties to reach the required ignition temperature $T\geq T_{n}$
and explode when struck by a WIMP.  We will consider a variety of WIMP masses,  
$m_{\chi}=10,100\ \text{and\ }1000$
GeV.

\section{Temperature Increase}

We proceed now to calculate the temperature increase given to a nanoparticle
by a WIMP collision with a nucleus in the metal. While Eq.[\ref{eq.heat_capacity}] gives
a rough idea of the temperature increase, here we will compute this quantity more carefully.
A metal nucleus recoiling from a WIMP interaction moves a certain distance before stopping. The amount of energy per length lost by the metal nucleus as it traverses through the metal nanoparticle and/or oxide is given by the stopping power ($S_{i,j}=-\frac{dE}{dx}$). Consequently, the range (stopping distance) is given by
\begin{equation}
r_{f}=\int^{E_0}_0 \frac{dx}{dE}dE=\int^{E_0}_0\frac{dE}{S_{i,j}},
\end{equation} 
where $E_0$ is the initial energy of the metal nucleus and $S_{i,j}$ is the total stopping power given by Lindhard (1961) theory \cite{Lindhard} when considering a nucleus of type $i$ moving through a medium of type $j$. Lindhard theory is valid for heavy ions with a few keV of energy. A different theory, Bethe-Bloch, needs to be used when considering backgrounds like $\alpha$ or $\beta$ particles, since they have larger energies (usually a couple of MeVs) and a small charge ($Z=1$ or $2$). The stopping power of backgrounds will be discussed in a later section.

The total stopping power for a slowly moving heavy ion can be separated into the electronic $S^e_{i,j}$ and nuclear $S^{\nu}_{i,j}$ components, such that $S_{i,j}=S^{\nu}_{i,j}+S^e_{i,j}$. Utilizing Lindhard's equations \cite{Lindhard}, we can approximate both stopping powers as:
\begin{eqnarray}\label{stopping}
S^{\nu}_{i,j}=2.8\times10^{-15}n_j\frac{Z_{i}Z_{j}}{(Z_{i}^{2/3}+Z_{j}^{2/3})^{1/2}}\frac{M_{i}}{M_{i}+M_{j}}(eV-cm^{2}),\\
S^e_{i,j}=1.2\times 10^{-16}n_jZ_i^{1/6}\frac{Z_iZ_j}{(Z_i^{2/3}+Z_j^{2/3})^{3/2}}\sqrt{\frac{E}{M_i}}(eV^{1/2}-cm^2).\label{stopping1}
\end{eqnarray} 
In the equations for the nuclear and electronic stopping power, the subscripts $\{i,j\}$ correspond to the parameters of the scattered nucleus and the medium respectively, $Z$ is the atomic number, $n_j$ is the atomic number density of the medium and $M$ is the atomic mass measured in atomic mass units. 
 
In this case $S_{1,1}$ pertains to a metal nucleus moving through the
metal nanoparticle and $S_{1,2}$ is the stopping power of a metal nucleus
moving through the outside medium. We will neglect any geometric factors and assume that all of the energy of the recoiling
nucleus will be deposited into the nanoparticle if $r_{f}\leq2R_{n}$.
If $r_{f}>2R_{n}$ then only a fraction of the
recoiling energy will be deposited into the nanoparticle. In the latter case, when the ion escapes the metal nanoparticle, the energy deposited into the nanoparticle $\Delta E=E_i-E_f$ can be found by solving for the final energy $E_f$ in the following equation:
\begin{equation}
2R_{n}=\int^{E_0}_{E_f}\frac{dE}{S_{1,1}}.
\end{equation}
The effective amount of energy deposited into the nanoparticle,
$E_{eff}$, is given by
\begin{equation}
E_{eff}=\begin{cases}
E_0 & \text{if }r_{f}\leq2R_{n}\\
\Delta E & \text{otherwise.} 
\end{cases}
\label{effectiveenergy}
\end{equation}
It is this quantity that determines the temperature increase of the metal nanoparticle.  We take the entire nanoparticle to be
heated uniformly by this amount of energy.  The initial temperature increase of the nanoparticle is then given by a modified version  
of Eq.(\ref{eq.heat_capacity}),
\begin{equation}
\label{eq:heat_increase}
\Delta T_{0} =  E_{eff}/C_{n} .
\end{equation}
Let us assume that the detector is operating at some uniform background temperature $T_{un}$. For example,
it might be at room temperature $T_R$.  Then the entire nanoparticle is initially heated by the WIMP interaction to 
\begin{equation}
T_0 = T_{un} + \Delta T_{0}.
\end{equation}

 In time, the temperature increase of the nanoparticle due to the WIMP interaction dissipates.  
 We will now solve the heat transfer equation in order to follow the evolution of the temperature profile $T(r,t)$.  
 This allows us to calculate the temperature near a single nanoparticle as a function of time in order to establish the characteristic time scale for how long the metal particle remains heated, and give us its temperature at the
 time of dissipation.  The goal is to determine whether the temperature found at the interface is greater than the ignition temperature assumed for our model
$T_{n}\geq 3,800$ K, in which case the nano-thermite in question could  plausibly  work as a WIMP detectors. 

The heat transfer equation is solved for a single sphere of radius $R_{n}$ embedded into a semi-infinite medium with different thermal properties. The
heat diffusion of a sphere to a surrounding medium has been analytically
calculated by A. Brown (1965) \cite{Brown}. Our calculation assumes zero thermal
resistance at the interface. Thus, the calculated temperature increase may be much lower than in the real system, since a finite resistance will contain
the heat inside the nanoparticle. In the following calculation, we will study the temperature increase $\Delta T(r,t)=T(r,t)-T_{un}$ profile as a function of radial distance ($r$) and time ($t$). Specifically, the heat transfer equation
\begin{eqnarray*}
\partial_{t}T_{1} & = & \alpha_{1}\nabla^{2}T_{1}\ \ \text{for}\ r\leq R_{n}\\
\partial_{t}T_2 & = & \alpha_{2}\nabla^{2}T_{2}\ \ \text{for}\ r>R_{n}
\end{eqnarray*}
is solved given the thermal conductivity ($k_i$) and thermal diffusivity ($\alpha_i$) of the sphere and medium. The subscript ``1" pertains to the sphere, whereas the subscript ``2" is given to the medium parameters. The following initial conditions (IC) and boundary conditions (BC) are chosen in order to solve the heat transfer equation:
\begin{eqnarray}\label{refbc}
T_1= T_0-T_{un}\equiv \Delta T_{0}\ \text{and \ }T_{2}=0\ \text{at\ } t=0\notag \\
T_1=\text{finite}\ \text{at \ } r=0 \\
T_1=T_2\ \text{and \ }k_{1}\frac{dT_{1}}{dr}=k_{2}\frac{dT_{2}}{dr} \ \text{at \ }r=R_{n}.\notag
\end{eqnarray}
The boundary condition found in the last line implies that there is no thermal resistance and no heat source at the interface.

\begin{figure}[h]
 \centering
\includegraphics[width=1.1\linewidth]{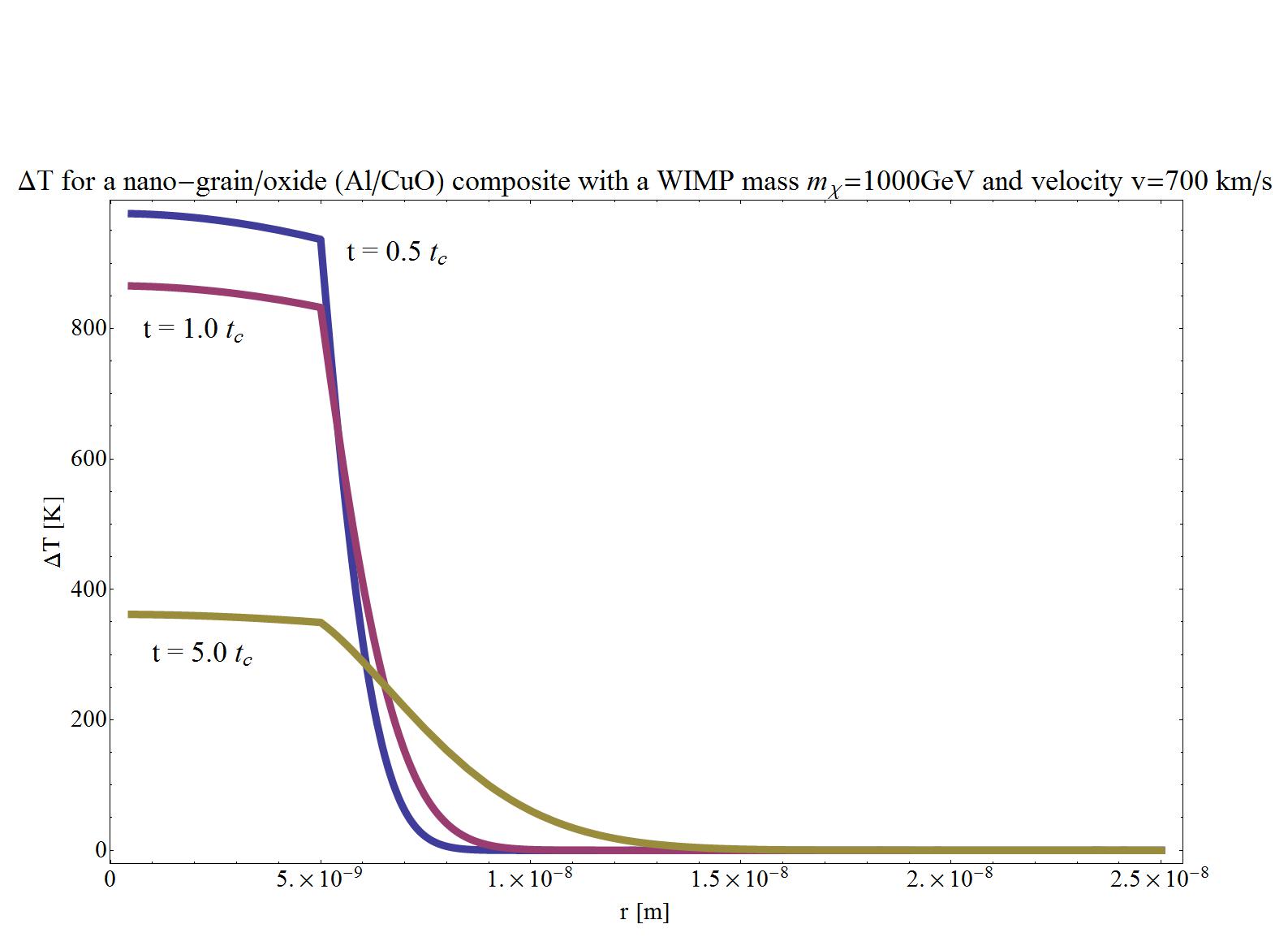}
    \caption{This figure shows the change in temperature $\Delta T$ after a WIMP/nucleus collision with maximum energy $E_{max}$ and speed of $v=700$ km/s. The different plots represent $\Delta T$ at times $t = 0.5\ t_{c}$, $t = t_{c}$ and $t = 5\ t_{c}$ as a function of the distance from the center of the metal nanoparticle. The nanoparticle significantly cools shortly after the conduction time, $t_{c}=\frac{R_{n}^2}{\alpha}$. This feature is general to all materials considered. The rapid cooling at times greater than $t_{c}$ is due to the exponential term in Eq [\ref{DeltTt}]. Thus, $t_{c}$ serves as a very good estimate for the total time the nanoparticle is heated. }
    \label{temphist2}
\end{figure}

Following (Brown 1965)\cite{Brown}, we find that the temperature is given by

\begin{eqnarray}
\frac{T_{1}}{\Delta T_{0}}(r,t)=\frac{2QR_{n}}{\pi r}\int_{0}^{\infty}\frac{(\sin u-u\cos u)\sin\left(\frac{ur}{b}\right)\exp(-u^{2}\frac{\alpha_{1}}{R_{n}^{2}}t)}{(u\cos u+L\sin u)^{2}+(Qu\sin u)^{2}}du\\
\frac{T_{2}}{\Delta T_{0}}(r,t)=\frac{2R_{n}}{\pi r}\int_{0}^{\infty}\frac{(\sin u-u\cos u)F(u)\exp(-u^{2}\frac{\alpha_{1}}{R_{n}^{2}}t)}{(u\cos u+L\sin u)^{2}+(Qu\sin u)^{2}}\frac{du}{u},
\end{eqnarray}
where
\begin{eqnarray}
F(u)=(u\cos u+L\sin u)\sin\left(\frac{u(r-R_{n})}{\sigma R_{n}}\right)+Qu\sin u\cos\left(\frac{u(r-R_{n})}{\sigma R_{n}}\right)\\
\sigma=\sqrt{\frac{\alpha_{2}}{\alpha_{1}}}\ \ \ \ \ \ \ \ \ \ \ Q=\frac{k_{2}}{k_{1}}\sigma\ \ \ \ \ \ \ \ \ \ \ L=\frac{k_{2}-k_{1}}{k_{1}} .
\end{eqnarray}
It can be seen from the solution above that there is a natural time-scale $t_{c}$ that arises from solving the heat eaquation; $t_{c}=\frac{R_{n}^2}{\alpha _1}$. This  time-scale, which will be referred to as conduction time, gives a modest estimate as to how long does the metal nanoparticle stay heated.
Figure [\ref{temphist2}] shows how the temperature diffuses from the nanoparticle to the medium at  times $t={0.5t_{c},t_{c},5t_{c}}$. Thus, one can approximate the amount of time that the nanoparticle stays at a high temperature by $t_{c}$. The temperature found at the
 interface $(r=R_{n})$
measured at a time $t=t_{c}$ is the relevant temperature value that needs to be compared to the ignition temperature $T_n$ in order to determine if the thermite reaction will ignite. The temperature increase at the interface (r=$R_{n}$) is given by: 

\begin{eqnarray}
\Delta T(R_{n},t) & = & T(R_{n},t)-T_{un}\\
 & = & \frac{2Q}{\pi}\frac{\alpha_{1}E_{ff}}{k_{1}}\int_{0}^{\infty}du\frac{\sin(u)(\sin(u)-u\cos(u))\exp(-u^2\frac{\alpha_1}{R^2_{nano}}t)}{(u\cos(u)+L\sin(u))^{2}+(Qu\sin(u))^{2}}\label{DeltTt}
\end{eqnarray}
so that 
\begin{eqnarray}
\Delta T(R_{n},t_{c})=\frac{2Q}{\pi}\frac{\alpha_{1}E_{eff}}{k_{1}}\int_{0}^{\infty}\frac{\sin(u)(\sin(u)-u\cos(u))\exp(-u^{2})}{(u\cos(u)+L\sin(u))^{2}+(Qu\sin(u))^{2}}du.\label{eq:t*}
\end{eqnarray}
In the previous equations for $\Delta T$, we have explicitly made the substitution $\Delta T_{0}=\frac{E_{eff}}{c_1\rho_1V}$ and definition $\Delta T(R_n,t)\equiv T_1(R_n,t)=T_2(R_n,t)$.

As can be easily appreciated, there are many physical parameters that
can influence the temperature. 
To reduce the number of parameters we need to study, we will make the following simplifying assumptions.

We will use the semi-empirical mass formula to approximately relate the atomic
number to the atomic mass: 
\begin{equation}\label{Z}
Z_i=0.5\frac{\frac{M_i}{\text{amu}}}{1+7.7\times10^{-3}\left(\frac{M_i}{\text{amu}}\right)^{2/3}}.
\end{equation}
Thus we eliminate $Z_1$ as a free parameter in the calculation of the of the stopping power in Eq (\ref{stopping}-\ref{stopping1}), and consequently the effective energy in Eq (\ref{effectiveenergy}).

The thermal diffusivity will be simplified by taking advantage of
the relation $\alpha=\frac{k}{c\rho}$ and the Dulong-Petit Law which
states that the heat capacity of a solid in crystalline form is given
by $c=24.9\frac{\text{amu}}{M}$ $\frac{\text{J}}{\text{gK}}.$ Thus, the thermal diffusivity
is approximated by
\begin{equation}
\alpha_{i}=\frac{k_{i}}{\rho_{i}}\left(\frac{1}{24.9\frac{\text{amu}}{M_i}}\right).\label{eq:nnano}
\end{equation}
It should be noted that the Dulong-Petit Law overestimates the heat
capacity for light atoms bonded strongly to each other at room temperature
such as beryllium, and for most solids kept at cryogenic temperatures
(i.e. 4.2 K or 77 K). In both cases, the overestimation of the heat capacity
will give a smaller thermal diffusivity, which as a consequence gives
a smaller $\Delta T$ in Eq $[\ref{eq:t*}]$. Thus, use of the Dulong-Petit is a conservative assumption that leads to a calculated temperature increase smaller than reality. 

Taking into account Eq \ref{Z} and Eq \ref{eq:nnano}, the number of parameters is reduced to five:
density ($\rho$), thermal conductivity ($k$) and atomic mass ($M$) of the metal, 
the WIMP velocity ($v$), and the WIMP mass
 ($m_{\chi}$).   Now that we can calculate the temperature increase as a function of
the 5 chosen parameters ($\rho,\ k,\ M,\ v,\ m_{\chi}$), we can study
which metal would produce a high enough temperature to overcome
the ignition temperature needed. Note, that the calculation for the temperature increase is independent of the assumed ignition temperature $T_{n}$. Thus, we can select the optimal metals from our calculations of the temperature increase, and perform future experiments to confirm the viability of the nano-thermite as a dark matter detector.  

\section{Results}

The equation for the temperature increase gives the temperature profile
for a sphere of a given material surrounded by a medium of another
material. Specifically, Eq {[}$\ref{eq:t*}${]} will be used in order
to calculate the temperature increase expected for a metal sphere
embedded into a specific oxide; copper (II)-oxide. 
Thermite reactions work by putting two metals of very different reactivity together and 
letting the more reactive (in a chemical sense) steal the oxygen from the least reactive.  
Specifically, we choose copper (II) oxide (CuO) as the material for the medium 
due to the low reactivity of copper. Then any metal more chemically reactive than
copper will create a thermite reaction when exposed to CuO at a temperature
greater than the ignition temperature $T_n$. The fact that copper is one
of the least reactive metals allows us to consider a large family of
metals to use as nanoparticles. Another very promising metal-oxide is tungsten trioxide (WO$_3$). Like copper, tungsten is also a very low reactive metal. The study of which metal-oxide would work best is beyond the scope of this paper and warrants further examination. 

\begin{figure}[!h]
\centering
\includegraphics[width=0.496\linewidth]{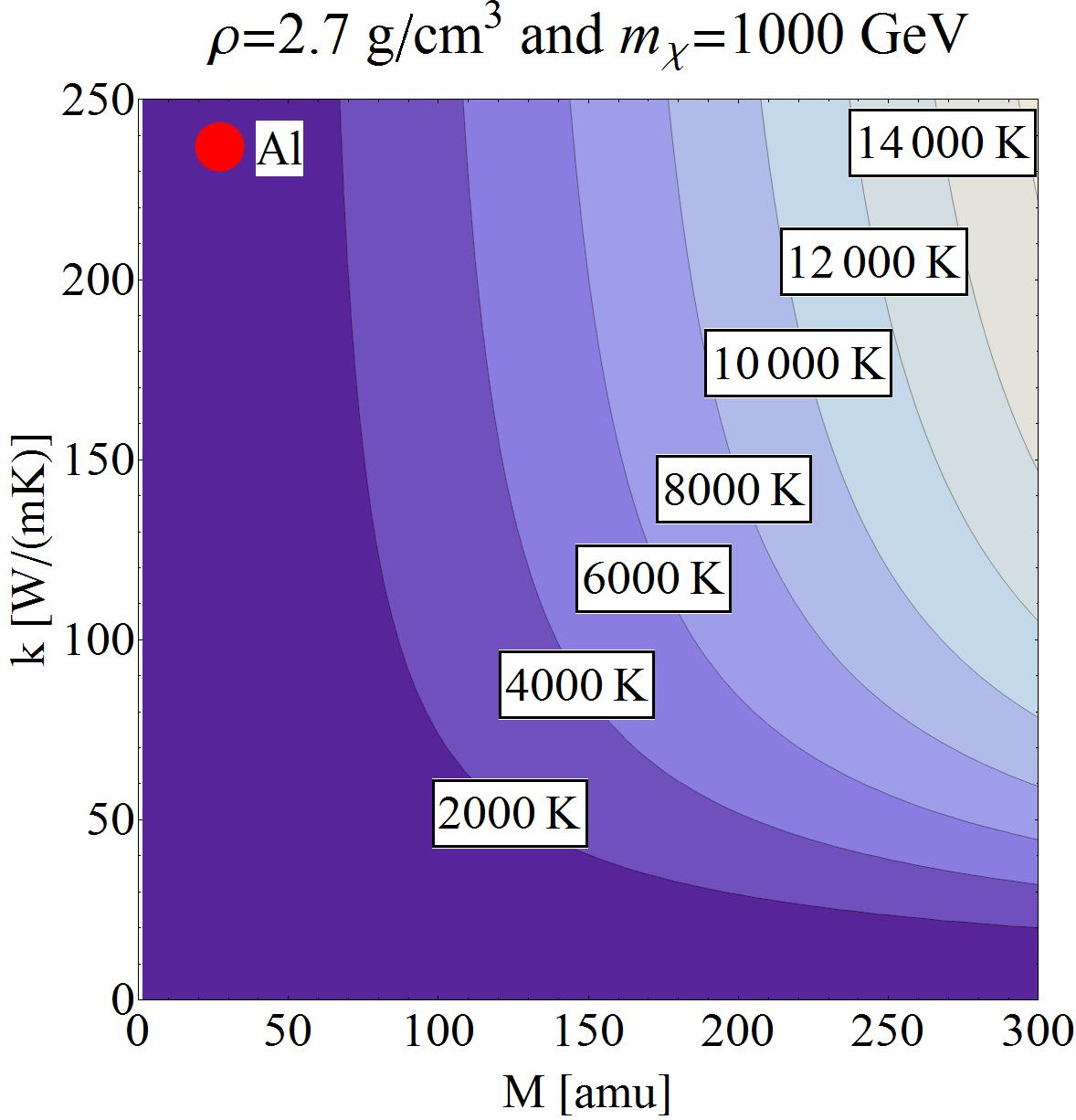}
\includegraphics[width=0.496\linewidth]{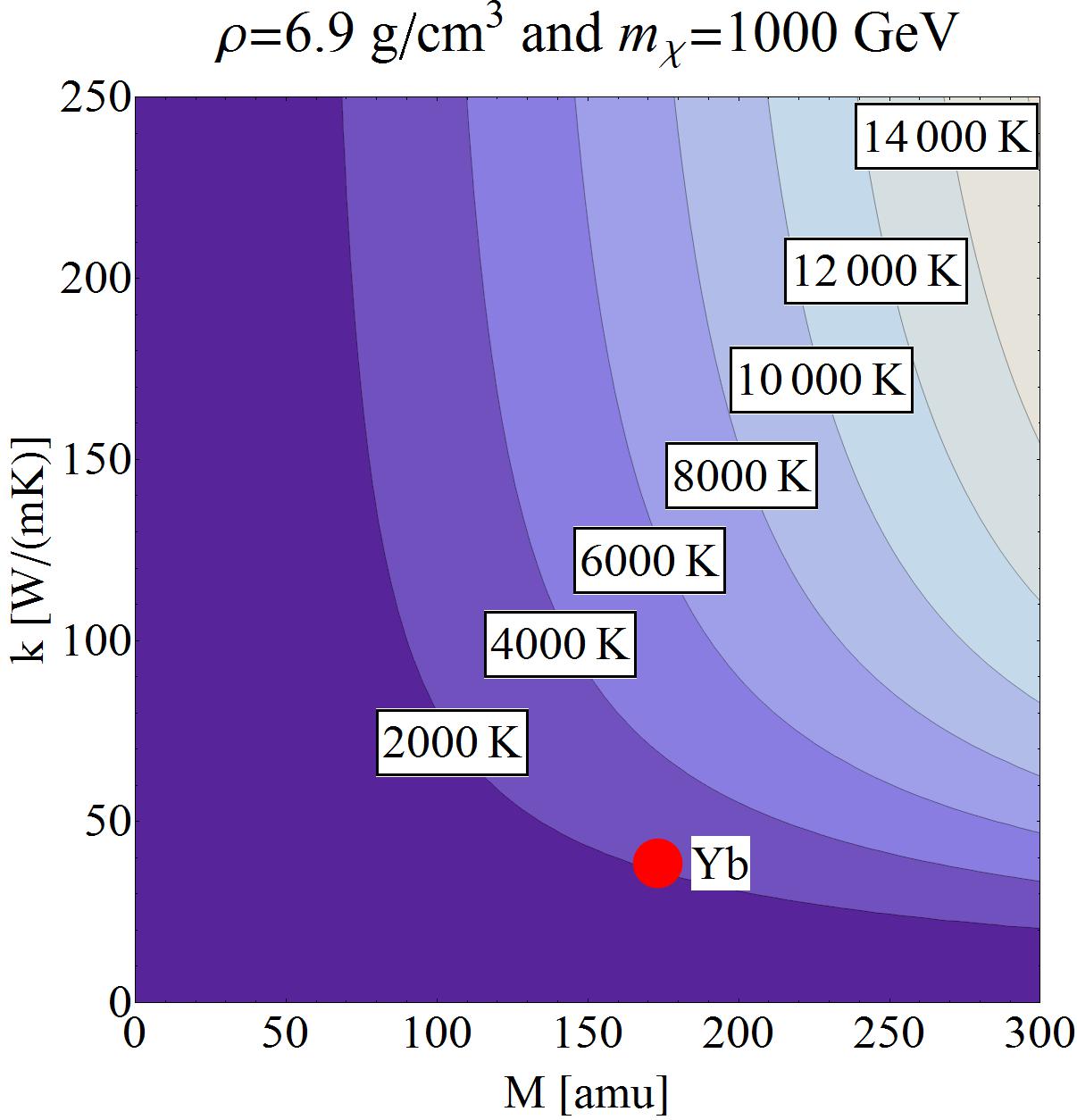}
\includegraphics[width=0.496\linewidth]{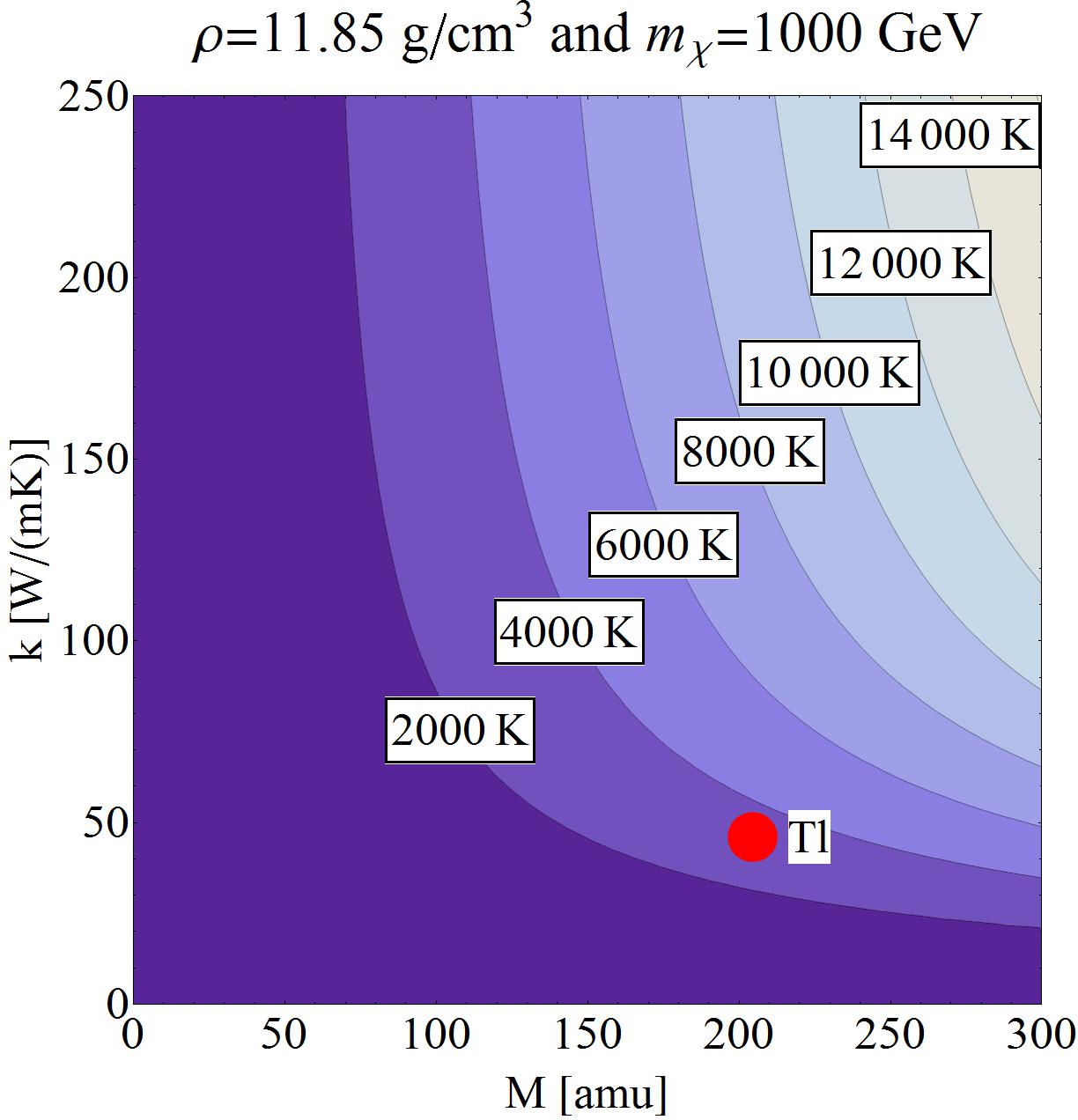}
\includegraphics[width=0.496\linewidth]{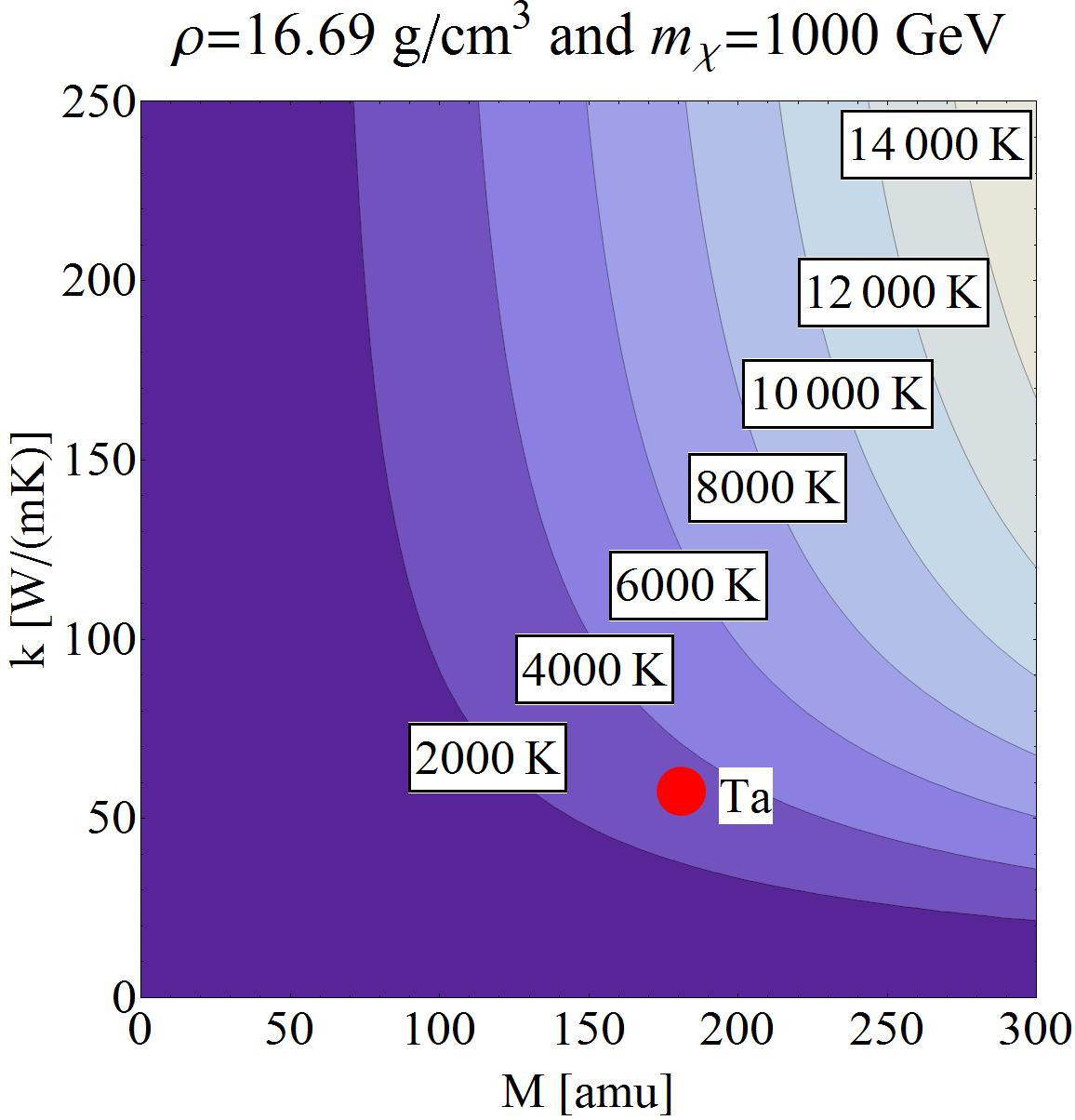}
\caption{These contour plots shows the temperature increase $\Delta T$ as a function of atomic mass $M$ and thermal conductivity $k$ of the metal used, given the densities:  $\rho=2.7\ \text{(Al)},\ 6.9\ \text{(Yb)},\ 11.85\ \text{(Tl)},\ \text{and}\ 16.69\ \text{(Ta)}$ g/cm$^3$. The chosen densities correspond to the metals: Aluminum, Ytterbium, Thallium and Tantalum, respectively. The red dots on each contour plot shows where each metal lies. The reason for showing these graphs, even though they span an unreal parameter space of density, thermal conductivity and atomic mass, is to show the general trend of the temperature output as a function of {$\rho,\ k\ \text{and}\ M$}. The hope is that the reader can familiarize himself/herself with this trend and, if interested, possibly explore a new set of metals as potential candidates for a nano-thermite WIMP detector.}
\label{contour}
\end{figure}

We wish to calculate the temperature
increase caused by a WIMP/metal-nucleus elastic collision at the interface
of the metal nanoparticle with copper (II) oxide for a couple of metal
targets in order to identify potential metals that could produce
enough temperature to create a thermite reaction. Four different metals
are chosen as test metals: aluminum, ytterbium, thallium and tantalum.
Figure [\ref{contour}] shows the reason for this choice of metals. The
first argument is the fact that their densities range
from one of the lightest elements in the periodic table, aluminum,
to one of the densest active metals, tantalum.

In order to study which metals would work best for the nanoboom detector
under different limits of the WIMP/nucleus interaction, the analysis
is divided into two distinct sections.
The two differing cases are: I) typical recoil energy, $\frac{E_{max}}{2},$
is deposited to the metal nucleus at the mean speed for galactic WIMPs
in the lab frame ($v=300$ km/s), and II) a very energetic WIMP with
a velocity close to the escape velocity ($v=700$ km/s) deposits the
maximum energy , $E_{max},$ to the metal nucleus. For each case, the
temperature increase at the interface is calculated for each of the
four metals and assuming a WIMP mass of either $m_{\chi}=10\ \text{GeV},\ 100\ \text{GeV},\ 1000\ \text{GeV}$.
Given a mass for the WIMP and a specific consideration of the recoil
energy, either case (I) or (II), then the temperature increase $\Delta T$
calculated for the metal nanoparticle is dependent on three further parameters:
the density ($\rho)$ and thermal conductivity $(k$) of the metal
and the mass of the metal nucleus $(M).$ The plot shown in Figure [\ref{contour}] gives $\Delta T$
assuming a WIMP mass of $m_{\chi}=1000$ GeV, a recoil energy $\frac{E_{max}}{2}$
with speed $v=300$ km/s and four different densities: $\rho=2.7,\ 6.9,\ 11.85\ \text{and }16.69$ $\text{g/cm}^{3},$
which are the densities of aluminum, ytterbium, thallium and tantalum
respectively. The reason for choosing Ytterbium, Thallium
and Tantalum is due to their relatively large mass ($m_{Yb}=173.05$
amu, $m_{Tl}=204.35$ amu and $m_{Ta}=180.95$ amu), which gives a
higher temperature increase. Figure [\ref{contour}] shows that for a $1000$ GeV
WIMP $\Delta T$ increases for metals with a higher thermal conductivity
and mass. The fact that a large thermal conductivity produces a higher
temperature is seen clearly in the boundary condition. Looking at
Equation \ref{refbc} one can see that if $k_{1}\gg k_{2}$
then $\frac{dT_{1}}{dr}|_{r=b}\rightarrow0$, which is the boundary
condition found for an insulator. The difference in thermal conductivity between the metal and the oxide induces a phonon spectrum mismatch at the interface, which creates an effective thermal resistance. This implies that materials with a very large thermal conductivity compared to CuO will retain the heat inside, and consequently have a higher temperature at the interface.   
The recoil energy grows as the difference between
the mass of the metal nucleus and WIMP mass decrease, because the term $\frac{\mu^2}{M}$ in the equation for $E_{max}$ is maximized for $M=m_\chi$. Thus, the temperature
output produced by the WIMP/metal-nucleus interaction will increase
as the mass of the metal increases for very heavy WIMPs ($m_\chi \geq 300$ GeV) and gets closer to the WIMP mass. In contrast, the temperature increase for low mass WIMPs (i.e. $m_{\chi}=10$ GeV) will be greatest for metals with low atomic mass (e.g. aluminum), closest to the mass of the WIMP.  

Aluminum is a low atomic mass metal with a high thermal conductivity, which are helpful characteristics for the study of low mass WIMPs. Furthermore, Al$^{27}$ is an attractive target for the study of spin-dependent WIMP/nucleus interactions. As well, aluminum is a popular metal fuel in thermite reactions. Aluminum based thermite reactions
have been heavily studied within the scientific community and used
in industry for a long time. By considering aluminum as one of our
metal targets, we gain perspective as to what type of thermite reaction
should be studied in order to make the nanoboom detector work. 

\begin{table}
\begin{tabular}{|c|c|c|c|c|}
\hline 
Element Name & Aluminum & Ytterbium & Thallium & Tantalum\tabularnewline
\hline 
\hline 
$\rho$ {[}$\text{g/cm}^{3}${]} & 2.7 & 6.9 & 11.85 & 16.69\tabularnewline
\hline 
$M$ {[}amu{]} & 26.98 & 173.05 & 204.38 & 180.95\tabularnewline
\hline 
$k$ {[}W/(mK){]} & 237 & 38.5 & 46.1 & 57.5\tabularnewline
\hline 
$\Delta T_{300}(m_{\chi}=10\ GeV)$ {[}K{]} & 190 & 37 & 26 & 22\tabularnewline
\hline 
$\Delta T_{300}(m_{\chi}=100\ GeV)$ {[}K{]} & 483 & 1,605* & 1,260* & 992\tabularnewline
\hline 
$\Delta T_{300}(m_{\chi}=1000\ GeV)$ {[}K{]} & 510 & 2,155* & 3,222* & 3,272\tabularnewline
\hline 
$\Delta T_{700}(m_{\chi}=10\ GeV)$ {[}K{]} & 504 & 407 & 288 & 245\tabularnewline
\hline 
$\Delta T_{700}(m_{\chi}=100\ GeV)$ {[}K{]} & 749 & 2,282* & 3,360* & 3,465*\tabularnewline
\hline 
$\Delta T_{700}(m_{\chi}=1000\ GeV)$ {[}K{]} & 832 & 2,767* & 4,078* & 4,221*\tabularnewline
\hline
\end{tabular}
\caption{This table gives the temperature increase for the four chosen metals when interacting with WIMPs of differing mass with a recoil energy given by one of the two considered cases. The temperature is given in kelvins. The subscript $300$ ($700$) of $\Delta T$ specifies if the recoil
energy considered was  from Case I (Case II). For comparison, the superscript ``*" indicates a temperature increase higher than the melting temperature of its respective metal. The total temperature at which the metal heats up is given by adding room temperature to the resulting temperature increase: $T_{total}=\Delta T + 300 $ K. The total temperature $T_{total}$ is then compared to $T_{ig}=3.8\times 10^4$ K in order to establish which metals would work as nano-thermite detectors at room temperature. Only thallium and tantalum could work at room temperature. Nevertheless, aluminum and ytterbium are possible metals for a nano-thermite detector operating at cryogenic temperatures.} 
\end{table}
The table shows the results for the total temperature increase
found for the chosen four metals when interacting with WIMPs of differing
mass with a recoil energy given by one of the two cases considered. The relevant parameters of copper (II) oxide needed to calculate the temperature increase are: density ($\rho=6.315$ g/cm$^3$), thermal conductivity ($k =17$ W/(mK)) and heat capacity ($c=0.53$ J/(Kg))\cite{conductivity} \cite{heatcapacity}.
The subscript $300$ or $700$ of $\Delta T$ specifies if the recoil
energy considered was a typical elastic collision with $E_{nr}=\frac{E_{max}}{2}$
at a speed of $300$ km/s, or if the collision was very energetic
with $E_{nr}=E_{max}$ at a speed close to the escape velocity, $v=700$
km/s, respectively. For comparison, the superscript ``*" indicates a temperature increase higher than the melting temperature of its respective metal. The total temperature at which the metal heats up is given by adding room temperature to the resulting temperature increase: $T_{total}=\Delta T + 300 $ K. In order to establish which metals would work as nano-thermite WIMP detectors, we then compare the total temperature $T_{total}$ to the ignition temperature $T_{ig}=3.8\times 10^4$ K.

The results show that ytterbium has the highest temperature increase of the four metals when considering Case I and WIMPs of mass $m_\chi = 100$ GeV. Ytterbium increases to a temperature of $\Delta T_{300}=1,605$ K whenever a typical collision happens between a WIMP of mass $m_\chi = 100$ GeV (i.e. Case I). Ytterbium is the best metal out of the studied four in Case I for a WIMP of mass $m_\chi = 100$ GeV, because the term $\mu^2/M$ in the equation for $E_{max}$ Eq [\ref{eq:Emax}] is maximized for ytterbium. Note that tantalum and thallium have a higher temperature when considering Case II and a WIMP mass $m_{\chi}\geq 100$ GeV, because they are much denser metals than ytterbium. Thus, the stopping distance of a metal nucleus moving through thallium and tantalum would be shorter than in ytterbium. This implies that thallium and tantalum recoiling nuclei would deposit a higher fraction of its total energy into the nanoparticle.  
\\
\indent The calculations show that a tantalum metal nanoparticle at room temperature could get hotter than $ T=\Delta T + 300\  \text{K}\geq 3.8 \times 10^3$ K when considering highly energetic collisions ($E_{nr}=E_{max},$
$m_{\chi}\geq100$ GeV and $v=700$ km/s). Thus, metal nanoparticles composed of
tantalum could serve as a WIMP detector operating at room temperature
for $m_{\chi}\geq 100$ GeV. Similarly, a nano-thermite composed of thallium metal nanoparticles could detect WIMPs with mass $m_{\chi}\geq 1000$ GeV. In contrast, the table also shows that aluminum
has a consistent temperature increase for a larger range of WIMP mass.
Specifically, if aluminum were to be a metal with an ignition temperature
$T_{n}^{Al}\geq 190$ K, then aluminum nanoparticles could work to measure
light WIMPs. Aluminum has a higher temperature increase for low mass WIMPs compared to ytterbium, thallium and tantalum, because it has the highest thermal conductivity and lowest mass difference to a 10 GeV WIMP. According to Eq [$\ref{eq:t*}$], Case II WIMP/metal elastic collision can produce $\Delta T\geq 190$ K for WIMPs with mass $m_\chi \geq 3$ GeV. It should be noted that a detector composed of Al/CuO and with an ignition temperature of $T_{n}^{Al}=190$ K has an energy threshold of about 2keV. Such a detector can be made stable if it operates at cryogenic temperatures.

\subsection{Cryogenic Detector}

Operating the detector at cryogenic temperatures could increase its sensitivity in measuring low mass WIMPs. We will consider configuring the detector at two different cryogenic temperatures: $T_{R}=77\ \text{K\ and\ }4.2$ K.  Lowering the temperature at which the detector works allows us to
consider thermite reaction with more favorable activation and ignition temperatures. As well, a cryogenic detector opens the possibility of using halide-based thermites or other decomposition reactions. Herein, we discuss the possibility of an oxide-based nano-thermite dark matter detector working at cryogenic temperatures.

Thus, following the same argument used to get equation Eq$[\ref{eq:Ta}]$
and Eq$[\ref{eq:Ti}]$, we would need a much less increase in temperature
in order to start the thermite reaction. Considering temperatures
of 77 K and 4.2 K as the original temperature for the detector, then
the ignition temperature will be $T_{n}^{77}\geq 980$ K and $T_{n}^{4.2}\geq51$ K
respectively. This is specially helpful for detectors of WIMP mass
10 GeV, which had a lower temperature increase. This variant on 
the original design is very promising, since it affects equally any
material used as the metal and allows us to work with a larger family
of thermite reactions. It is important to note that a cryogenic detector with a very low activation temperature would make the detector unstable at room temperature. Thus, the materials would also need to be synthesized at cryogenic temperatures. Neverthless, a thermite detector working at cryogenic temperatures
will be stable even when considering very low activation temperatures;
on the order of $T_{a}^{77}\geq7,900$ K and $T_{a}^{4.2}\geq411$ K for
detector temperatures of $77$ K (liquid nitrogen) and $4.2$ K (liquid helium) respectively. Specifically, an Al/CuO nano-thermite detector with an ignition temperature of $T_{n}^{4.2}\geq 51$ K permits an energy threshold of 0.5 keV.  
Only experiments would determine the actual ignition temperature of an Al/CuO nanothermite. Fortunately, the ignition temperature of thermites is proportional to the size of the nanoparticle (i.e. a smaller nanoparticle has a lower ignition temperature), and other similar metals/oxide to Al/CuO could be used in order to achieve a ignition temperature close to 50K\cite{Davin}. The latter consideration is beyond the scope of this paper.  

\subsection{Background}
\indent Any WIMP detector is vulnerable to background coming from the cosmic rays or radiation from its components. Cosmic Rays can be minimized by putting the detector deep inside a mine underground. Nevertheless, background will still be present due to radiation decay inside the detector, which are mostly due to natural impurities of the materials. The energy deposed per unit length of the background onto the detector can be calculated by using Bethe-Bloch theory. The Bethe-Bloch equation, in contrast to the Lindhard equation, is used for relativistic ionizing particles like alphas and betas. The stopping power of alphas and betas can be found utilizing the ASTAR and ESTAR programs [\cite{ASTAR},\cite{ESTAR}]. Since the stopping power is linearly proportional to the density of the medium, we calculate the stopping power for alphas and betas moving in the densest material (Platinum) common to both programs. 

Excellent background rejection can be achieved because of the nanoscale granularity of the 
 detector. Single charged particles have a very long range. Alpha particles, for example, have a much longer range in the detector than do recoiling nuclei, which can be used to get rid of background due to alphas.  
 The energy loss of particles moving through the detector cells is $dE/dx \sim Z^2/\beta^2$. For helium $Z^2$ = 4 (near the end of the range the ion will pick up electrons and  dE/dx increases, but for the most part one can take $Z=2$ for helium).  Z is much higher for nuclear recoils due to WIMP interactions, causing faster energy loss;
 thus the nuclei stop within a single cell. The WIMP makes only one cell explode:
 the chain reaction initiated by one exploding metal nanoparticle is restricted to nanoparticles within only one cell, which is thermally isolated from neighboring cells by an insulating material.
 On the other hand, the range of the $\alpha$-particles, usually on the order of $10\  \mu$m or greater, is longer than size of one cell, and therefore make about 20 cells explode at once.  Thus
 the nanoscale granularity is key for background rejection.  

 Alpha particles produced in a radioactive decay usually have energies around 5 MeV. Utilizing ASTAR, the stopping power of an alpha particle moving through platinum with an energy of 5 MeV is $S_\alpha=5.02$ keV/(10 nm). Alphas have a large enough stopping power to produce a signal in a nano-thermite detector composed of a highly dense metal and energy threshold $E_{th}<5 $ keV. Fortunately, the range of alpha particles is much larger than that of heavy ions. Rejection of alpha particle signals can be made due to the nanometer granularity of our detector. Specifically, the typical range of a heavy ion is around 50 nm; whereas, an alpha particle moving through a material would typically travel approximately $10\ \mu $m or more. As an illustrative example, the stopping distance for a 5 MeV alpha particle moving through Al$_2$O$_3$ (SiO$_2$) is 13.76 $\mu$m (20 $\mu$m). Since each cell has a size of $0.5\ \mu $m, then a heavy ion would only explode 1 cell; and an alpha particle would explode 20 or more cells. Signals produced by alpha particles can be rejected by eliminating all exploded cell clusters.  
 
The energy range of beta particles is wide, ranging from a few keV to hundreds of MeVs. This occurs, because the energy of the radioactive decay is usually shared between an electron and a neutrino. Specifically, the biggest stopping power for a beta particle moving through platinum in the range of energies $10\ \text{keV} < E < 100 $ MeV(found at energy $E=100 $ MeV) is $S_\beta=0.33$ keV/(10 nm). A nano-thermite detector will produce a signal if one or more nanoparticles ignite and initiate a chain reaction within the cell. Thus, if our detector has an energy threshold greater than $E_{th}>0.5 $ keV, then beta particles will not produce a signal in our detector. It should be noted that the range of beta particles is much longer than alpha particles. Thus, assuming that a beta particle can deposit enough energy to the nanoparticle to make an explosion and/or the detector has a lower energy threshold of $E_{th}<0.5$ keV; then the same method of rejecting signals originated by alpha particles could be used to identify explosions caused by beta particles. The signal produced by a beta particle (if any) would explode many cells, much more than alpha particles. This result would hold true also if we considered other metals with a lighter density, since the quoted stopping power is from one of the densest metals in the periodic table. Another source of background are gammas, which are highly energetic photons. 

Even though gamma particles are electrically neutral, they can create photo electrons through the photoelectric effect and the Compton effect. Either of those interactions will eject an electron at relativistic speeds, turning it into a beta particle that will ionize many more atoms. Typical energies for gamma particles produced in a radioactive decay range from a few hundred keV to 10 MeV. ESTAR can be employed to learn the behavior of the electrons produced by the interaction of an atom with a gamma ray. By the same argument found for beta particles above, gammas will not produce a signal in a nano-thermite WIMP detector with energy threshold $E_{th}>0.5$ keV. The nanometer granularity of the nano-thermite detector is such that background from beta and gamma particles do not produce any signal (assuming an $E_{th}>0.5$ keV), and signals originated by alpha particles can be discarded due to the large number of cells exploded.

\section{\label{sec:Summary} Summary}

We have studied the ignition properties of nanoscale explosives as a novel type of dark matter detector. Other design concepts may be employed for the nanothermite dark matter detector, which could obtain lower energy thresholds and/or measure directionality of the recoiling nucleus sourced by a WIMP/nucleus interaction.
We focused on two-component nanothermite explosives consisting of
 a metal and an oxide. As a specific example, we considered metal nanoparticles of 5 nm radius embedded in a gel of oxide, with millions of these nanoparticles constituting one ``cell" isolated from other cells. A large number of cells adds up to a total of 1 kg detector mass. A WIMP striking a metal nucleus in the nanoparticle, deposits energy that may be enough to initiate a reaction at the interface between the two layers. 
 
We calculated the temperature increase of a metal nanoparticle due to a WIMP interaction and compared
it to the ignition temperature required for the nanoparticle to explode.  We computed the range of the nuclear recoil using the Lindhard formula;
if the recoiling nucleus did not stop inside the nanoparticle, we considered only the fraction of the energy that was deposited inside the metal nanoparticle itself.  
This energy fraction was then converted to a temperature increase.  We needed to know how long the nanoparticle remained hot in order to 
determine whether an explosion was set off.  This timescale was obtained from the heat transfer equation. All assumptions made during the calculations were chosen in the spirit of being as conservative as possible. 
We then compared the temperature increase to the ignition temperature
required to set off a nanothermite explosion.  This ignition temperature varies for different thermite materials, and was computed by
requiring two conditions to be met:
  (i) for each of the thermites we considered, there should be no spontaneous combustion of any of the metal nanoparticles for at least a time period of one year, 
  and (ii) the temperature increase from a WIMP interaction must be sufficiently high to overcome an activation barrier
and allow the thermite reaction  to proceed.  
We searched through a variety of thermite materials to find those whose temperature increases from WIMP interactions would 
exceed their ignition temperatures for an explosion.  We found aluminum, ytterbium, thallium and tantalum to be particularly suited to discover WIMPs via the explosion they would induce. We note that our model assumed that both the metal and oxide interact as solids. However, if the metal changes physical state into a gas due to a correspondingly high temperature increase, then the nano-thermite reaction rate may drastically increase. 

Excellent background rejection can be achieved because of the nanoscale granularity of the 
 detector. The WIMP makes only one cell explode:
 the chain reaction initiated by one exploding metal nanoparticle is restricted to nanoparticles within only one cell, which is thermally isolated from neighboring cells by an insulating material.
 The range of the $\alpha$-particles on the other hand is longer than size of one cell, on the order of $10\ \mu$m, and therefore makes approximately 20 or more cells explode at once.  Thus
 the nanoscale granularity is key for background rejection.  Betas and gammas, on the other hand, rarely set off an explosion at all.

We found a number of thermites that would serve as efficient WIMP detectors.  Using a single model, we found that
if the detector operates at room temperature,  WIMPs with masses
above 100 GeV (or for some materials above 1 TeV) could be detected; they
 deposit enough energy ($>$10 keV) to cause  an explosion.  
When operating cryogenically at liquid nitrogen or liquid helium temperatures, 
the nano explosive WIMP detector can detect energy deposits as low as 0.5 keV, making the nano explosive detector
sensitive to very light  $<$10 GeV WIMPs. Even with the conservative model presented in this paper, our calculations suggest that oxide-based nano-thermites would work as a dark matter detector. We look forward to experiments which will establish accurately the minimal energy deposition by a recoiling nucleus necessary for a nano-thermite combustion.  
\\


\begin{acknowledgments}
  K.F.\ and A. L.\ acknowledge the support of the DoE under grant DOE-FG02-95ER40899 and the Michigan
  Center for Theoretical Physics.  A.K.D., K.F., and A.L.\ acknowledge the MCubed grant via the University of Michigan for support.
  
\end{acknowledgments}







\end{document}